# Photoinduced Dirac semimetal in $ZrTe_5$

T. Konstantinova[1], L. Wu[1], W.-G. Yin[1,*], J. Tao[1], G. D. Gu[1], X. J. Wang[2], Jie Yang[2], I. A. Zaliznyak[1,*], and Y. Zhu[1,*]

[1] *Brookhaven National Laboratory, Upton, NY 11973, USA*

[2] *SLAC National Accelerator Laboratory, Menlo Park, CA 94024, USA*

**Novel phases of matter with unique properties that emerge from quantum and topological protection present an important thrust of modern research. Of particular interest is to engineer these phases on demand using ultrafast external stimuli, such as photoexcitation, which offers prospects of their integration into future devices compatible with optical communication and information technology. Here, we use MeV Ultrafast Electron Diffraction (UED) to show how a transient three-dimensional (3D) Dirac semimetal state can be induced by a femtosecond laser pulse in a topological insulator $ZrTe_5$. We observe marked changes in Bragg diffraction, which are characteristic of bond distortions in the photoinduced state. Using the atomic positions refined from the UED, we perform density functional theory (DFT) analysis of the electronic band structure. Our results reveal that the equilibrium state of $ZrTe_5$ is a topological insulator with a small band gap of ~ 25 meV, consistent with angle-resolved photoemission (ARPES) experiments. However, the gap is closed in the presence of strong spin-orbit coupling (SOC) in the photoinduced transient state, where massless Dirac fermions emerge in the chiral band structure. The time scale of the relaxation dynamics to the transient Dirac semimetal state is remarkably long, $\tau \sim 160$ ps, which is two orders of magnitude longer than the conventional phonon-driven structural relaxation. The long relaxation is consistent with the vanishing density of states in Dirac spectrum and slow spin-repolarization of the SOC-controlled band structure accompanying the emergence of Dirac fermions.**

*zaliznyak@bnl.gov, wyin@bnl.gov, zhu@bnl.gov

## Introduction

The strong interest in topological Dirac and Weyl semimetals is rooted both in their fundamental attraction as model systems for experimenting with theories of particle physics and in their unique electronic properties, such as the suppressed backscattering, peculiar surface states, chiral and spin-polarized transport, and novel responses to applied electric and magnetic fields controlled by topological invariance, which are promising for technological applications [1, 2, 3, 4]. Recently, significant attention gained theoretical ideas of how to prepare these phases on demand by photoexcitation and periodic driving by external stimuli (Floquet state engineering) [5, 6, 7, 8]. Much less explored are the pathways towards experimental realizations of these theoretical proposals. The first strides in this direction are made by the very recent results on $WTe_2$ and $MoTe_2$, where metastable crystal symmetry change leading to a topologically distinct phase was induced by THz light pulses [9, 10]. Here, we discover a transition to a topologically distinct electronic phase which does not rely on the change of the macroscopic symmetry of the crystal lattice and therefore can be photoinduced without an accompanying crystallographic phase transition. This discovery is significant because it provides a pathway towards tuning the band structure topology that is decoupled from complexities, domain size limitations, and relaxation phenomena associated with a bulk phase transition between different crystalline phases[11].

Zirconium pentatelluride, $ZrTe_5$, is a remarkable topological material. It was theoretically predicted to be at the phase boundary between a weak topological insulator (TI) and a strong topological insulator [12] and initially was proposed as a rare example of 3D Dirac semimetal (DSM) [13, 14, 15]. Recent high-resolution ARPES [16, 17, 18] and transport [19] measurements suggest that $ZrTe_5$ is a weak topological insulator, but with a very small band gap (~ 20 meV) at the $\Gamma$ ($q = 0$) point and a tiny Fermi surface hosting charge carriers with very small mass [14, 15]. The WTI nature was corroborated by observing the topological edge states at the steps of $ZrTe_5$ crystalline surface [20]. The bandgap of the insulating state closes under pressure and at 6.2 GPa $ZrTe_5$ becomes a (possibly topological) superconductor [21]. Recently, evidence for strain-tuned topological phase transition from weak to strong topological insulator with the Dirac semimetal state at the critical point was reported in $ZrTe_5$ [22]. Here, we discover that a topological phase transition from TI to DSM can be induced transiently, by ultrafast optical pumping.

An indication that ultrafast optical control of the topological electronic band structure in $ZrTe_5$ is possible has recently been obtained in surface-sensitive time-resolved (tr)-ARPES experiments, suggesting potential for technological applications in ultrafast optoelectronics [23]. A downshift of the valence energy band was observed following photoexcitation with a sub-picosecond, 1.55 eV photon pulse, which was attributed to transient heating after optical perturbation. The corresponding electronic equilibration times in the valence and conduction bands, 1.6 ps and 0.8 ps, respectively, were consistent with the conventional electron-lattice relaxation [24, 25]. Similarly short, few-ps phonon-mediated relaxation times were observed for hot carrier dynamics in $ZrTe_5$ from time-resolved optical reflectivity using similar optical pumping, with even shorter time (~ 0.25 ps) for electron-electron thermalization [26]. Here, we use bulk-sensitive MeV UED to study the temporal dynamics of the photoinduced structural changes in $ZrTe_5$. We discover a transient metastable atomic structure with a markedly longer thermalization time scale (~ 100-200 ps) consistent with SOC-controlled relaxation [27] and deduce the corresponding electronic band structure as the DSM state using DFT calculations.

**Results**

**Experimental procedure and results**

$ZrTe_5$ is a layered material, where each layer is formed by quasi-one-dimensional trigonal prismatic $ZrTe_3$ chains with non-equivalent apical (Te1) and dimer (Te2) tellurium atoms running along the crystalline $a$-axis, which are bound together along the $c$-axis via pairs of Te (Te3) atoms forming zigzag chains [Fig. 1(a)]. The prismatic chains and the zigzag Te3 chains form two-dimensional sheets of $ZrTe_5$ in the $ac$ plane, stacking along the $b$-axis via weak van der Waals interaction. Nevertheless, electronically $ZrTe_5$ is a 3D material, with a closed, albeit strongly anisotropic Fermi surface ($k_{\mathrm{F}}^{c}/k_{\mathrm{F}}^{a} \sim k_{\mathrm{F}}^{c}/k_{\mathrm{F}}^{b} \sim 5$, where $k_{\mathrm{F}} = (k_{\mathrm{F}}^{a}, k_{\mathrm{F}}^{b}, k_{\mathrm{F}}^{c})$ is Fermi wavevector) and strongly anisotropic dispersion and electron mass [3, 15, 16, 17, 28]. We use the orthorhombic lattice notation [space group Cmcm (#63)] with lattice parameters $a = 3.9943$Å, $b = 14.547$ Å, and $c = 13.749$ Å at 300 K and two formula units per unit cell [12, 29] for the steady state of $ZrTe_5$. The atomic Wyckoff positions are shown in Supplementary Table 1.

In our work, thin single crystals of $ZrTe_5$ exfoliated from high quality bulk crystal samples [3, 28] are excited by 1.55 eV, 60 fs laser pulses with the polarization in the *ac* crystal plane and probed via diffraction of 100 fs, 4.0 MeV electron pulses. Optical properties of the material in the photon energy region around 1.55 eV are obtained with ellipsometry measurements. The imaginary part, $\langle\varepsilon_2\rangle$, of the pseudo dielectric function, $\langle\varepsilon\rangle = \langle\varepsilon_1\rangle + i\langle\varepsilon_2\rangle$, at 300 K is shown in Supplementary Figure 1; there is weak anisotropy of $\langle\varepsilon_2\rangle$ in this range. The 1.55eV energy is close to the inter-band transition, which is centered around 1.3 eV. Inter-band transitions can lead [25] to lattice deformation if the intensity of the excitation is sufficiently high. Experiments were performed at temperatures of 55 K and 27 K, with an incident excitation fluence of 3.5 mJ/cm$^2$. No significant difference in dynamical lattice behavior at these two temperatures was observed.

A typical UED diffraction pattern obtained prior to the photoexcitation is shown in Fig.1(b) and the difference pattern after and before the photoexcitation in Fig.1(c). Here, the diffraction intensity measured at a large time-delay, $t \approx 800$ ps (averaged within the time window $[707, 1000]$ ps), where the lattice dynamics reach quasi-equilibrium, is subtracted from the intensity at a negative time delay, $t < 0$, before arrival of the pump pulse (averaged within the 160 ps time window). A remarkable change in Bragg reflections is observed, which is most clear for the $(00L)$ peaks with systematic increase in intensity for most of the reflections. The intensity of $(00L)$ Bragg reflections is sensitive to intra-unit-cell atomic displacements, $\delta z_\nu$, along the $c$-axis direction ($\nu$ indexes atom at position $z_\nu$ and with atomic scattering factor $f_\nu$). In the frozen-phonon model of a static distortion of the crystal lattice, such displacements lead to an intensity modulation of $(00L)$ peaks, $\Delta I_{00L} \sim \left|\sum_\nu f_\nu (L \cdot \delta z_\nu) \mathrm{e}^{-\mathrm{i}(2\pi L \cdot z_\nu)}\right|^2$ [30], which is roughly consistent with the observed photoinduced change of the Bragg diffraction pattern. These observations clearly indicate an involvement of the displacements of Te2 and Te3 atoms [Fig. 1(a)], which reside at low-symmetry 8f Wyckoff positions in the equilibrium $ZrTe_5$ structure [29] (Supplementary Table 1), with $z_{\mathrm{Te2}} = 0.151$ and $z_{\mathrm{Te3}} = 0.434$, respectively, along the $z$ ($c$-axis) direction. While the above formula for intensity change induced by displacement obtained in kinematic, single-scattering (Born) approximation is very useful for gaining qualitative understanding of atomic displacements, it can rarely (as in Ref. [9]), if at all, be used for the quantitative analysis of electron diffraction where the interaction of the scattering electrons with the system is strong and multiple scattering effects cannot be neglected. Here, we quantitatively analyze the lattice structure of $ZrTe_5$ and refine $z_{\mathrm{Te2}}$ and $z_{\mathrm{Te3}}$ positions from the UED patterns before and after photoexcitation using the Bloch-wave dynamical scattering theory, which accurately accounts for multiple scattering effects. The corresponding calculated electron diffraction pattern for the unperturbed equilibrium structure [29] is shown in Fig. 1(d).

## Temporal dynamics of scattering

An important further insight is provided by the quantitative analysis of the temporal dynamics of Bragg peak intensities encoding the photoinduced atomic displacements, which is presented in Fig. 2. At the early stage of the photoexcitation, the dynamics is reflected in intensity transfer from Bragg peaks to thermal diffuse scattering (TDS), Fig. 2(a). The decrease of Bragg peak intensities is due to the lattice deformation and the increased atomic vibration as a result of energy transfer from the photoexcited electrons to the lattice. The observed time constant that describes these initial dynamics of both TDS and total Bragg scattering is $\tau_1 \approx 3$ ps, which is consistent with phonon-phonon relaxation and is only slightly slower than the time scale of electron-phonon coupling measured by ARPES [23]. After the initial fast drop, the total Bragg scattering starts increasing, but on a very much slower time scale, $\tau_2 \approx 150$ ps. This increase is accompanied by a similarly slow further growth of diffuse scattering intensity, which indicates that both signals arise from the same underlying physics.

In our experiment, the intensity of the transmitted central beam is monitored by a separate detector

(see METHODS for details). This allows us to measure total scattering, which provides an important consistency validation. We notice that the intensity of the central beam has a downward trend within the 1 ns window, which becomes very clear after averaging over the multiple measured time-series, Fig. 2(b). The dynamics of the central beam can be described with a single exponent with time constant $\approx 150$ ps, which is consistent with the slow recovery of Bragg peaks and a concomitant increase in thermal diffuse scattering. This indicates that the depletion of the central beam is consequential to the increase of scattering governed by the slow relaxation process.

The individual Bragg peaks also reveal the same two-time-scales dynamics, with an initial fast drop governed by the time constant $\tau_1$ of few ps and a slow recovery within 1 ns window with a much longer relaxation time, $\tau_2 \sim 100 - 200$ ps. While intensity of the majority of Bragg peaks recovers after the initial drop, the level to which it recovers is strongly $q$-dependent [Fig. 2(c),(d)]. Some peaks, notably those having non-zero component along $(00L)$ direction, including $(00L)$ peaks in Fig. 2(c), reach higher intensity than in their initial, unperturbed state, which can be explained by the presence of atomic displacements along $z$ ($c$-axis) lattice direction in the transient metastable state. On the other hand, $(HH0)$ peaks show significantly different dynamics, with no recovery at all, Fig. 2(d). This indicates the absence of photoinduced atomic displacements in this direction, or a possibility that the relevant displacements are scattered in phase, yielding cancellation of their contributions to the observed Bragg intensity.

The two distinct relaxation times, $\tau_1$ and $\tau_2$, obtained by fitting the intensity of Bragg peaks at different wave vectors, $\mathbf{q}$, to two-time relaxation dynamics, $\Delta I = A\left(1 - e^{-t/\tau_1}\right) + B\left(1 - e^{-t/\tau_2}\right)$, are presented in Fig. 3(a). While the short relaxation time of a few ps, $\tau_1$, is consistent with the electron-phonon and phonon-phonon scattering mechanisms [24, 25], the long time, $\tau_2 = 160(30)$ ps, which is roughly two orders of magnitude larger, indicates that conventional electron-phonon scattering is suppressed and implicates a different mechanism, involving much weaker interactions compared to the Coulomb forces that govern electron and phonon scattering. This is entirely consistent with the involvement of SOC, which is broadly recognized to be important for determining the electronic properties of $ZrTe_5$ [3, 17, 18]. SOC is a relativistic interaction that in atomic systems has a degree of smallness described by the fine structure constant, $\alpha \approx 1/137$, i. e. is typically two orders of magnitude weaker compared to Coulomb interactions. SOC-controlled electronic spin relaxation times of the order of hundreds of picoseconds to few nanoseconds are indeed observed in semiconductor quantum wells [27,31, 32] and in bulk Ge [33] and graphene [34]. The long relaxation times result from the weakness of SOC compared to the Coulomb interaction that governs conventional charge relaxation, combined with the small density of conduction electrons. In the case of Dirac electrons in graphene, as well as in the present case of $ZrTe_5$, the latter is a consequence of the vanishing density of states near the Dirac point [35].

**Bloch-wave dynamical scattering structure refinement**

In order to quantitatively analyze the photoinduced changes of $ZrTe_5$ crystalline lattice and determine the positions of Te atoms in the transient metastable photoexcited state, we conducted a structural refinement by comparing the experimental electron diffraction patterns with the ones calculated using the Bloch-wave dynamical scattering method. We focus on the $(00L)$ reflections, of which sufficient number (up to $|L| = 20$) are present in our data and which show clear and systematic changes in the difference pattern [Fig. 1(c)]. As discussed above, this allows us to refine the positions, $z_{\mathrm{Te3}}$ and $z_{\mathrm{Te2}}$, of Te3 and Te2 atoms at the low-symmetry crystallographic 8f Wyckoff site and Debye-Waller (DW) factor, which quantifies the degree of crystallographic disorder and reflects the strength of TDS. We note that displacement of Te1 and Zr atoms along the $c$-axis alters the crystal symmetry of $ZrTe_5$ structure, which was not observed in our UED experiment.

The results of the refinement are presented in Figure 3(b), which shows the $(00L)$ Bragg intensity profiles measured by UED before (top) and after (bottom) the photoexcitation along with the best fits obtained using Bloch-wave ED calculations and varying the $z_{\mathrm{Te3}}$ and $z_{\mathrm{Te2}}$ atomic positions and DW factor. We used $\chi^2$ analysis (see METHODS) to evaluate the goodness of the fit. We first refine the diffraction pattern measured at the steady state, before the photoexcitation. In addition to the Te3 and Te2 atomic positions, in this refinement we also determine the sample geometry, including sample thickness and minor crystal tilt and bending, and DW factor, $B = 0.46$. The values of the atomic positions obtained in this refinement, $z_{\mathrm{Te2}} = 0.151$ and $z_{\mathrm{Te3}} = 0.434$, are in agreement with the results of Ref. [29] (Supplementary Table 1). We then keep the sample geometry unchanged and analyze the UED patterns in the photo-excited metastable state at long time delays, $t \approx 800$ ps. Here, we use an iterative procedure where we first keep Te3 atoms at their steady-state position and vary the Te2 position (green circles in Fig. 3(c)) to obtain $z_{\mathrm{Te2}} = 0.151(1)$ for the minimum $\chi^2 \approx 1.24$. We then vary the position of Te3 at this fixed position of Te2 [black circles in Fig, 3(c)] to determine $z_{\mathrm{Te3}} = 0.432(1)$ and $B = 0.52$, with significantly improved goodness of fit ($\chi^2 \approx 1.03$; for 430 free parameters in the fit, the 5% test of statistical significance corresponds to $\chi^2 \approx 1.11$). The increased value of DW factor is in agreement with the observed increase of TDS in the photoexcited state [Fig. 3(a),(b)]. We then again vary the Te2 atom position using the newly obtained $z_{\mathrm{Te3}} = 0.432$ [red circles in Fig.3(c)] to verify that the global minimum is achieved. We thus conclude that the observed marked changes of the Bragg reflection pattern in the transient photoexcited state induced by a femtosecond laser pulse result from a small but clearly identifiable displacement of the Te3 atoms and a much smaller, barely identifiable displacement of the Te2 atoms, which we were able to refine in our analysis.

**Discussion**

In order to understand the physics underlying the observed transient structural change, its relation to the electronic properties of $ZrTe_5$ and the role of SOC, we performed the first principles DFT calculations of the electronic band structure, which are presented in Fig. 4. The photoinduced structural modification in $ZrTe_5$ is characterized by the displacement of Te3 atoms in the 8f Wyckoff positions along the $c$-axis direction, from $z_{\mathrm{Te3}} = 0.434$ to $z_{\mathrm{Te3}} = 0.432$, which corresponds to an elongation of the Te3-Te3 bond by about 1% [Fig. 1(a) and inset of Fig. 3(a)]. Such a selective response of Te3 atoms to the laser pump agrees well with the electronic density of states (DOS) where the dominant contribution to the DOS at and above the Fermi level is provided by a large peak derived from 5p orbitals of the Te3 atoms [Fig. 4(a)]. The 1.55 eV incident photons excite electrons from the valence bands to the conduction bands, preferentially populating the Te3 derived states contributing to this peak. The resulting increased Coulomb repulsion pushes Te3 atoms apart, thus weakening the Te3-Te3 bond. Since the DOS at the Fermi level is predominantly contributed by the Te3 atoms' 5p orbitals, the seemingly small change in the structural parameters could affect the low-energy electronic structure significantly.

To explore the evolution of photoexcitation modified band structure of $ZrTe_5$, we performed a series of the first-principles calculations for $z_{\mathrm{Te3}} = 0.430 - 0.437$ (see Supplementary Discussion and Supplementary Figures 2-5 for details). The total energy minimum occurs at $z_{\mathrm{Te3}} = 0.434$ (Supplementary Figure 2), in good agreement with experiment. The corresponding calculated band structure reveals that the equilibrium state of $ZrTe_5$ has a small direct gap of $\approx$ 25 meV at the Γ-point and a smaller indirect gap between Γ and M points [Fig. 4(c)], also in good agreement with the ARPES results. In sharp contrast, at $z_{\mathrm{Te3}} = 0.432$ both gaps are closed and Dirac point emerges at the Brillouin zone center (Γ point), manifesting a photoinduced Lifshitz-type phase transition from topological insulator to Dirac semimetal [Fig. 4(d),(e)]. Wannier functional analysis of the electronic structure reveals a charge transfer of about 0.6% from Te1 to Te3 atoms accompanying this displacement (such displacement corresponds to the 15-19 meV $A_{\mathrm{g}}$ Raman-active modes of lattice vibrations [21]). This is consistent with the picture that the 1.55 eV laser illumination weakens the Te3-Te3 bond by selectively moving Te3 atoms apart. The calculated total energy per formula unit for the transient photoexcited state at $t \approx 800$ ps ($z_{Te3} = 0.432$) is only 12 meV higher than of the equilibrium structure at time zero ($z_{\mathrm{Te3}} = 0.434$). Remarkably, band structure calculations in the absence of SOC show that the gap is not closed, and the Dirac semimetal state is not formed neither at time zero, nor in the photoexcited structure at $t \approx 800$ ps [Supplementary Figure 4]. Hence, it is the strong SOC in the $ZrTe_5$ crystal that locks electronic spins and momenta, imparting electronic charge quasi-particles with chiral nature [3, 36] which is fully responsible for the formation of the transient photoinduced DSM state.

Strong SOC in $ZrTe_5$ has wide-ranging experimental consequences. In magnetotransport, SOC is manifested in a very large Lande factor, $g \approx 15.8 - 24.3$, observed from Zeeman splitting [13, 14, 37]. A circular dichroic (CD-)ARPES directly reveals the SOC-controlled chiral band structure, where electronic states near the Fermi level are spin-polarized [36]. This is consistent with the chiral magnetic effect observed in longitudinal magnetotransport [3]. Chirality protection can also explain the exceptional quasiparticle coherence, which is revealed by record-narrow ARPES spectra [17] and remarkably robust quantum transport underpinning a number of unusual and fascinating novel phenomena, such as vanishing quantum oscillations (zero-spin effect) [19], 3D quantum Hall effect [28], the anomalous Hall effect [38], and two-particle resonant states [39]. The exceptionally long, $160 \pm 30$ ps equilibration time of the photoexcited transient state in $ZrTe_5$, which DFT calculations show to be DSM observed in our present work, is very naturally explained by a spin-relaxation time of the SOC-governed chiral electronic structure. As we show in the Supplementary Figures 4 and 5, the closure of Dirac gap in $ZrTe_5$ band structure is governed by SOC. Hence, the newly emergent electronic Dirac quasiparticles are chiral (helical) [40], with spin-momentum locking that is distinct from that in the static TI phase. The electronic spin re-equilibration that is invariably required for the formation of the transient Dirac state relies on weak magnetic interactions and therefore must involve correspondingly a long SOC controlled spin relaxation time, such as observed in semiconductor quantum wells [27,31, 32], bulk Ge [33] and graphene [34], and as we observe here. The electron-phonon relaxation mechanisms, on the other hand, are suppressed due to the vanishing density of states in the Dirac electronic spectrum[35].

In conclusion, the simultaneously acquired temporal evolution of Bragg intensity of many reflections enables us to observe the subtle changes in atomic structure of complex electronic materials after pulsed laser excitation [11, 41, 42]. Here, we report a photoinduced deformation of the local structure in $ZrTe_5$, which selectively weakens specific Te-Te bond and underlies electronic topological phase transition to a Dirac semimetal. The optical excitation with a femtosecond laser pulse induces metastable transient state whose energy is only slightly above the ground state. The change in atomic positions in the photoinduced state is quantitatively determined from the Bragg intensities of UED through structural refinement using Bloch-wave dynamical scattering theory. The DFT band structure calculations show that in the presence of spin-orbit coupling this photoinduced meta-stable state features chiral Dirac quasiparticles, which emerge at the Γ-point. The equilibration time to the transient Dirac semimetal observed in this work is about two orders of magnitude longer than phonon-mediated thermalization times, consistent with the vanishing density of states in Dirac spectrum and slow spin re-polarization of SOC-governed chiral electronic state. We note that similar considerations might apply to equally long relaxation times of the photoexcited coherent atomic displacements observed in Bi, another well-known topological material with strong SOC [43]. Our results open exciting new opportunities for photon-induced band engineering of topological DSM states and call for a number of follow-up studies. Further support to our findings can be obtained by time-resolved CD-ARPES experiments [23, 36], which can directly probe the relaxation times of the chiral band structure. The coherent atomic displacements observed in our work indicate connection to a particular optic phonon mode, thus suggesting a possibility of energy-selective tuning of the band structure in $ZrTe_5$ by populating this mode using THz excitation.

When our manuscript was under review, we became aware of the optical pump-probe work by C. Vaswani, et al.[44], which fully corroborates our observations. There, a THz-pump-field induced metastable phase with unique, Raman phonon-assisted topological switching dynamics and with lifetime in excess of 100 ps was observed in $ZrTe_5$. Using first-principles modeling similar to ours, the authors identify a mode-selective Raman coupling driving the system from strong to weak topological insulator with a Dirac semimetal phase established at a critical atomic displacement. In their case, the topological transition is controlled by the phonon coherent pumping, as suggested above.

**Methods**

**Single crystalline** $ZrTe_5$ samples were prepared via a Te-flux method, as described in [3]. High purity Zr and Te elemental mixture $Zr_{0.0025}Te_{0.9975}$ were sealed under vacuum in a double-walled quartz ampule and first melted at 900 °C in a box furnace and fully rocked to achieve homogeneity for 72 h. The melt was followed by slow cooling and rapid heating treatment between 445 °C and 505 °C for 21 days, in order to re-melt crystals with small sizes. The resultant single crystals were typically about $0.1\times0.3\times20\text{mm}^3$. Crystals were chemically and structurally analyzed by powder x-ray diffraction, scanning electron microscopy with energy dispersive x-ray spectroscopy, and transmission electron microscopy. The samples for UED measurements were exfoliated using Scotch tape, yielding typical lateral crystal size around 50-100um. The flakes have an average thickness around 100 nm as determined with electron energy loss spectroscopy.

**Pump-probe diffraction experiments** were performed at SLAC using the 4.0MeV-UED facility with 3.5 $\text{mJ/cm}^2$ excitation fluence (1.55 eV pump) at 55 K and 27 K. The exposure time of a single shot for excitation and probing were 60 fs and 100 fs, respectively; 900 single-shot images were integrated at each time step. The time interval between measured points varied from 0.25 ps in the region of fast dynamics to 26 ps is the region of slow dynamics, as reflected by the density of points in Fig. 2. In the experimental setup the center beam was monitored by a separate detector. For quantitative intensity analysis and minimizing the effect of long-term electron count fluctuation due to the possible beam drifts intensities of the diffraction patterns in all the analysis were normalized by the corresponding value of the central beam intensity.

**Structure refinement** was carried out by quantitatively comparing the intensities of Bragg reflections before and after photoexcitation with the ones calculated using the Bloch-wave dynamical scattering method. We used $\chi^2 = \frac{1}{N}\sum_i \left(\frac{\boldsymbol{I}_i^{exp} - a - b\boldsymbol{I}_i^{cal}}{\sigma_i}\right)^2$ to evaluate the goodness of fit, where superscript "*exp*" refers to experimentally measured intensity and "*cal*" refers to calculated intensity, the subscript $i$ numbers data points in the $(0\ 0\ L)$ scan, and $N = 430$ is the total number of data points. $\sigma_i$ are the standard deviations, which were obtained as mean square deviations from the average for the set of diffraction patterns collected within the respective time window and $a$ and $b$ are fitting parameters corresponding to the background and intensity scaling constant. The calculated intensity accounts for diffraction from a single crystal flake with its plane horizontal and [010] parallel to the incident electron beam and a smaller twisted flake with [1-10] along the incident beam [see Fig. 1(c)]. The refinement of the diffraction patterns before photoexcitation was used to determine the sample geometry, which included sample thickness, small sample tilt and bending. From the fitting, the thickness of the [010] single crystal flake was determined to be 80 nm and the thickness of the smaller [1-10] flake was 51 nm. The minimum $\chi^2$ obtained in refinement before and after photoexcitation were 1.02 and 1.03, respectively.

**The first-principles calculations** were performed by using the WIEN2K (version 18.2) implementation [45] of the full potential linearized augmented plane-wave method in the generalized gradient approximation [46] of density functional theory with the SOC treated within the second variational method. The basis size was determined by $R_{mt}K_{max} = 7$ and the primitive Brillouin zone was sampled with a regular 24 × 24 × 6 mesh containing 628 irreducible k points to achieve energy convergence of 1 meV. The band structure was plotted in a dense 721 k-point path to show the opening and closing of the small gap at the Brillouin zone center.

**Data Availability statement**

The data that supports the findings of this study are available within the article [and its supplementary material].

**Acknowledgements**

We gratefully acknowledge discussions with Q. Li and J. Tranquada and help of T.N. Stanislavchuk with ellipsometry measurements. This work at Brookhaven National Laboratory was supported by Office of Basic Energy Sciences (BES), Division of Materials Sciences and Engineering, U.S. Department of Energy (DOE), under contract DE-SC0012704. We also acknowledge the use of Ellipsomentry facility at the Center for Functional Nanomaterials, a DOE user facility at BNL sponsored by the DOE BES Scientific User Facilities Division under the same contract. The SLAC MeV-UED is supported by the DOE BES Scientific User Facilities Division Accelerator and Detector R&D program under contract No. DE-AC02-76SF00515.

**Author Contributions**

Y.Z. conceived and directed the project. Y.Z. and T.K. coordinated the UED experiment with X.J.W. T.K. conducted the UED experiment with the assistance from J.Y. and J.T. G.G. provided the samples. T.K., L.W., and I.Z. analyzed the data and prepared the figures with input from Y.Z. L.W. carried out structure refinement using the Bloch-wave method. W.-G.Y. performed the ab initio DFT calculations. I.Z. and W.-G.Y. proposed the interpretation of the results. I.Z. and T.K. wrote the paper with input from Y.Z. and all authors.

**Competing Interests**

The authors declare no competing interests.

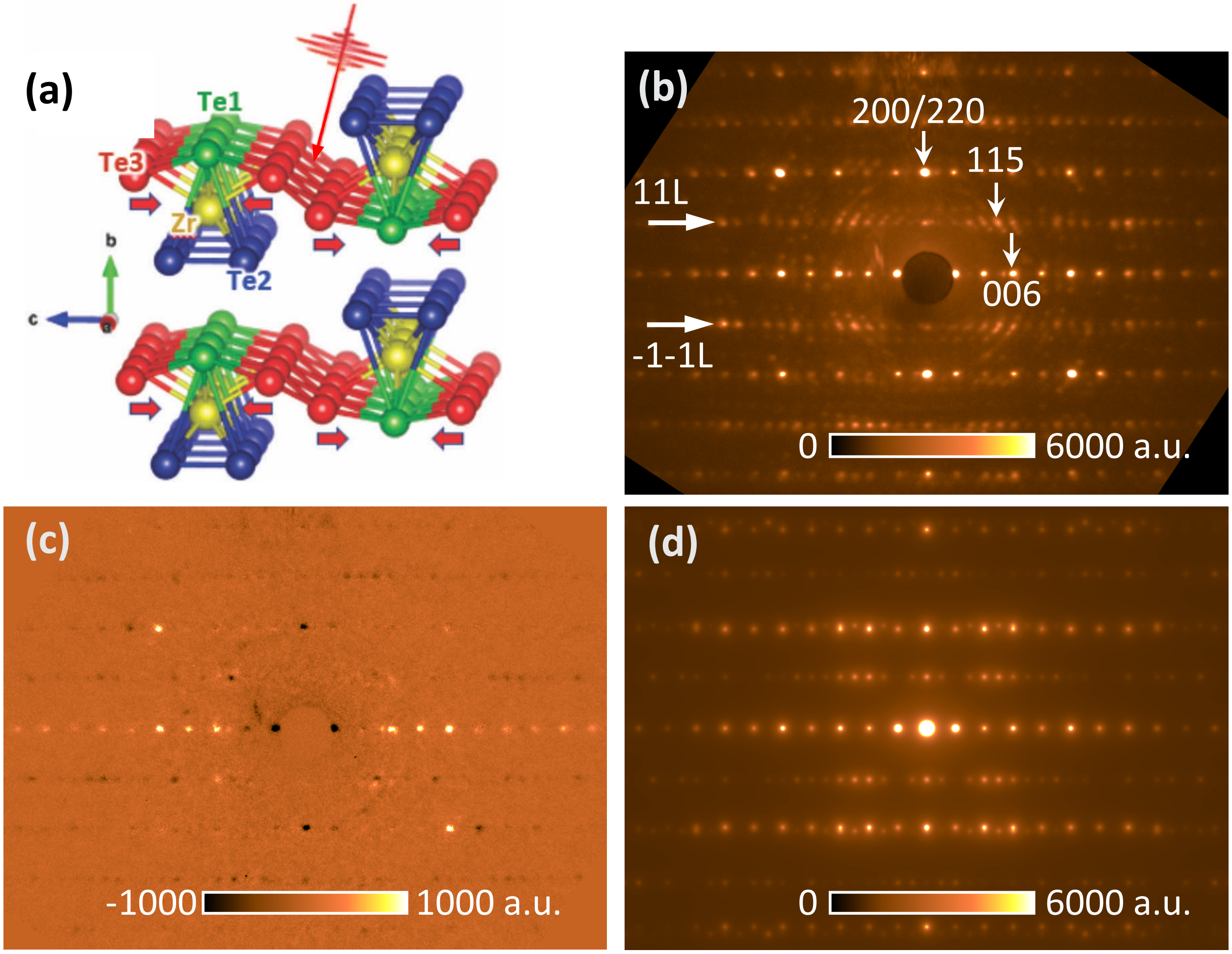

**Figure 1. Probing photo-induced dynamics of $ZrTe_5$.** (a) Schematics of the crystal structure. Different colors highlight the symmetrically non-equivalent apical (Te1, green), dimer (Te2, blue), and zigzag (Te3, red) tellurium atoms forming the $ZrTe_3$ chains and the $ZrTe_5$ sheets. Chains of $ZrTe_3$ prisms, consisting of Te1 and Te2 atoms extend along the a-axis and are connected by Te3 atoms along the $c$-axis to form $ZrTe_5$ layers in the $a-c$ plane. These layers are van der Waals bonded in the $b$-axis direction. Wiggly arrow illustrates selective photoexcitation of Te3 atoms; thick arrows mark the direction of atomic movement after photoexcitation. (b) UED pattern before photoexcitation. The diffraction is mainly from a single crystal flake with [010] as its plane normal and a smaller twisted flake with [1-10] normal, the latter contributes to the $(1\ 1\ L)$ and $(-1-1\ L)$ reflection rows as indicated by the horizontal arrows. (c) Difference pattern between before and after (at an $\approx$ 800 ps delay) photoexcitation. (d) Simulated diffraction pattern based on Bloch-wave method and using our structural refinement shown in Fig. 3 (a = 878, b = 2.5e8, see Methods), which agrees with the structure model from [29]. The calculation includes contributions of $(H\ 0\ L)$ zone with sample thickness of 80 nm and $(H\ H\ L)$ zone with sample thickness of 51 nm. The diffraction patterns are represented as color maps with color scales shown in the insets. The intensity is shown in arbitrary units corresponding to the readout of the CCD camera.

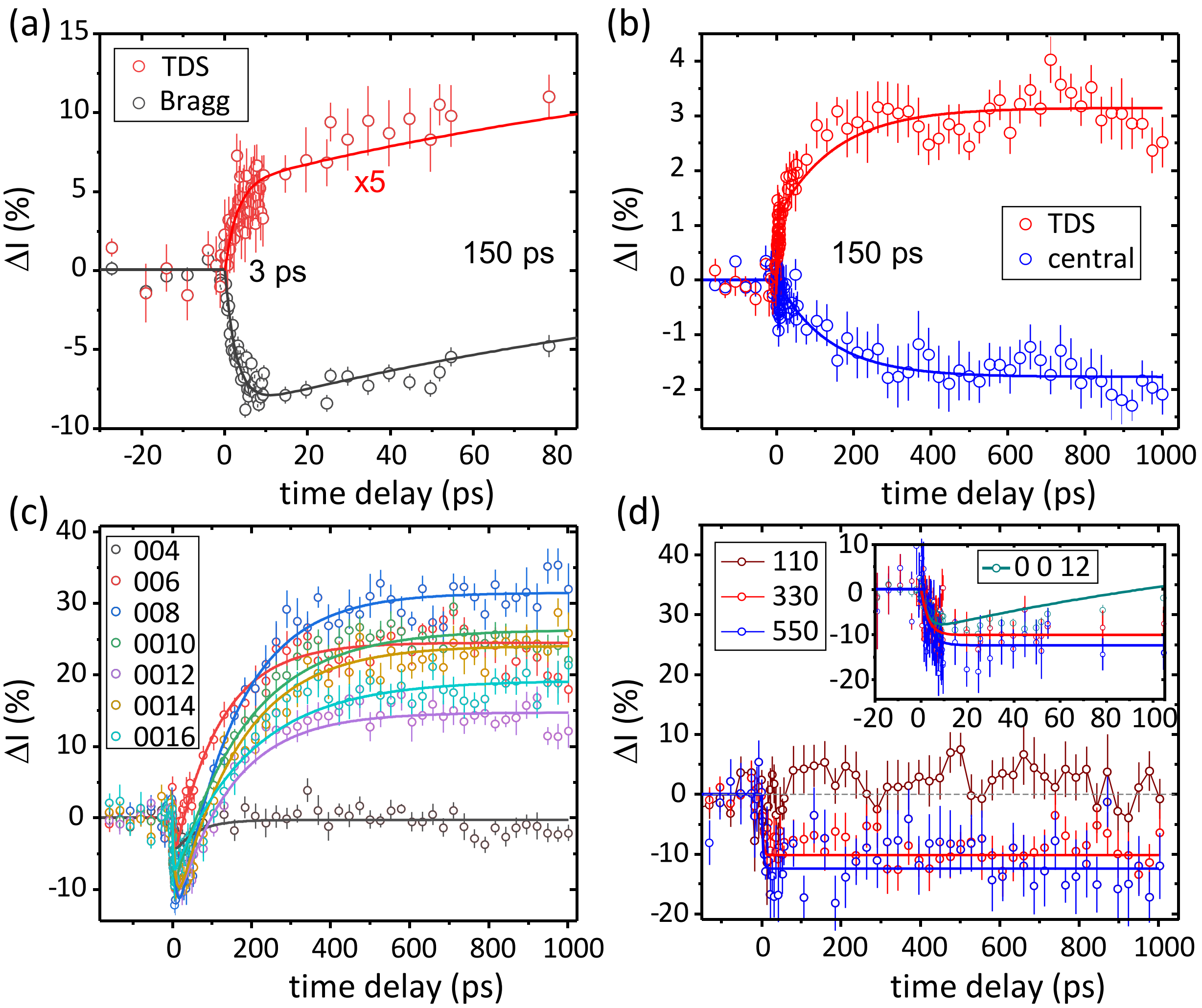

**Figure 2. Lattice dynamics of $ZrTe_5$ observed from photon pump and electron probe.** (a) Photo-induced dynamics of Bragg scattering averaged over (0 0 12), (0 0 14), (0 0 16), (3 3 5), (3 3 6), (6 0 0), (1 1 13), (1 1 14), (2 0 12), (2 0 16), (4 010) and their equivalent Bragg peaks (black circles) and that of thermal diffuse scattering (TDS) magnified by a factor of five (red circles) at short time delays, $\tau < 100$ ps. (b) Dynamics of TDS (red circles) and transmitted central beam (blue circles) at time delays up to 1 ns. Solid lines in (a) and (b) show fits to two-time exponential relaxation with $\tau_1 = 3$ ps and $\tau_2 = 150$ ps (see text). (c) Photo-induced intensity of the $(00L)$ peaks as a function of time delay. Open circles are experimental data and solid lines are fits to two-time relaxation describing the initial ($\tau_1 = 3 - 6$ ps) intensity drop via phonon mechanism, which is concurrent with the increasing TDS, and the long-time increase ($\tau_2 = 160(30)$ ps) associated with the photo-induced modification of the electronic band structure. (d) Lattice dynamics of the (odd odd 0) Bragg peaks, which are insensitive to the Te displacements associated with the electronic band reconstruction and only show the phonon-related short-time intensity drop. The inset zooms in at the initial drop of the intensity of these peaks and compares it with that of (0 0 12) peak at short time delays. The error bars in all panels show one standard deviation.

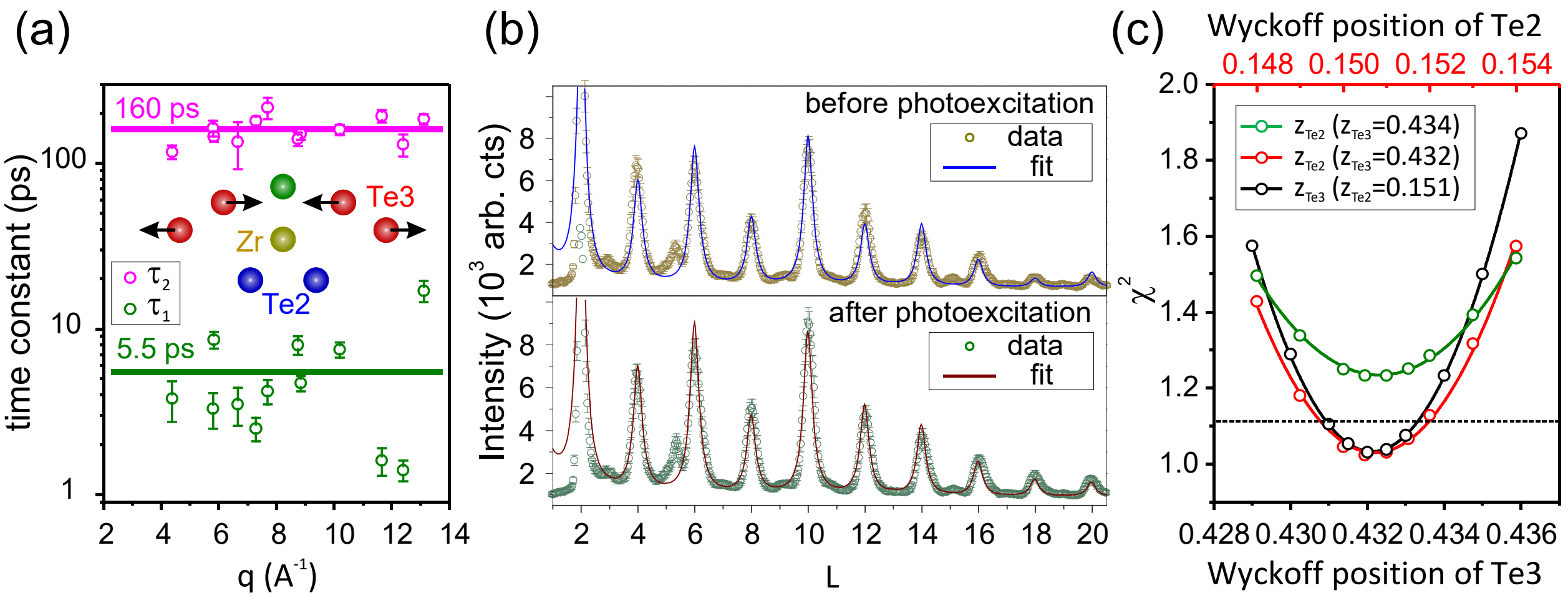

**Figure 3. Two distinct time scales of structural response and refinement of Te positions from UED.** (a) Two distinct time constants ($\tau_1$ and $\tau_2$) as a function of reciprocal vector, $q$, observed in $ZrTe_5$ in our pump-probe experiments. The thick solid lines show the average values of $\tau_1 = 5.5$ ps (green) and $\tau_2 = 160$ ps (magenta). The inset illustrates the structural model of $ZrTe_5$ with the arrows indicating the displacement direction of Te3 atoms after photoexcitation. (b) Measured UED intensity profiles (open circles) of the $(0\ 0\ L)$ reflections before (t < 0, top) and after (t > 700 ps, bottom) photoexcitation. The small peak on the left side of (0 0 6) is a contamination from the (1 1 4) reflection of the small flake and was discarded in the refinement. The error bars show one standard deviation. The solid blue line in the top panel is the calculated intensity profile based on the structure model in ref [29] and the DW factor $B = 0.46$ refined from our data. The red solid line in the bottom panel is calculated based on the same model but with $B = 0.52$ and $z_{\mathrm{Te3}} = 0.432$, both refined from the data. We first determine the sample geometry (thickness and bending) by fitting the intensity profile before photoexcitation (top) and then determine the displacement of $z_{\mathrm{Te2}}$ and $z_{\mathrm{Te3}}$ from the intensity profile after-photoexcitation (bottom) based on the goodness of fit ($\chi^2$). The (002) reflections, which are at the edge of the beam stopper cannot be reliably measured (the leading edge in both panels is visibly truncated) and are excluded from the refinement. (c) The value of $\chi^2$ quantifying the goodness of fit as a function of the Wyckoff position of Te3 and Te2 along $c$ direction for the intensity profile after-photoexcitation. The green and red circles show $\chi^2$ as a function of Te2 position with Te3 fixed at $z_{\mathrm{Te3}} = 0.434$ (green) and $z_{\mathrm{Te3}} = 0.432$ (red), respectively. The black circles show $\chi^2$ as a function of Te3 position with Te2 fixed at $z_{\mathrm{Te2}} = 0.151$. The solid lines are the second order polynomial fits used to determine the minimum $\chi^2$. The global minimum, $\chi^2 \approx 1.03$, is reached for $z_{Te2} = 0.151$ and $z_{\mathrm{Te3}} = 0.432$. The horizontal dotted line shows the level of $\chi^2 = 1.11$, which corresponds to 5% statistical significance test and determines the absolute accuracy of our refinements, $\pm 0.001$.

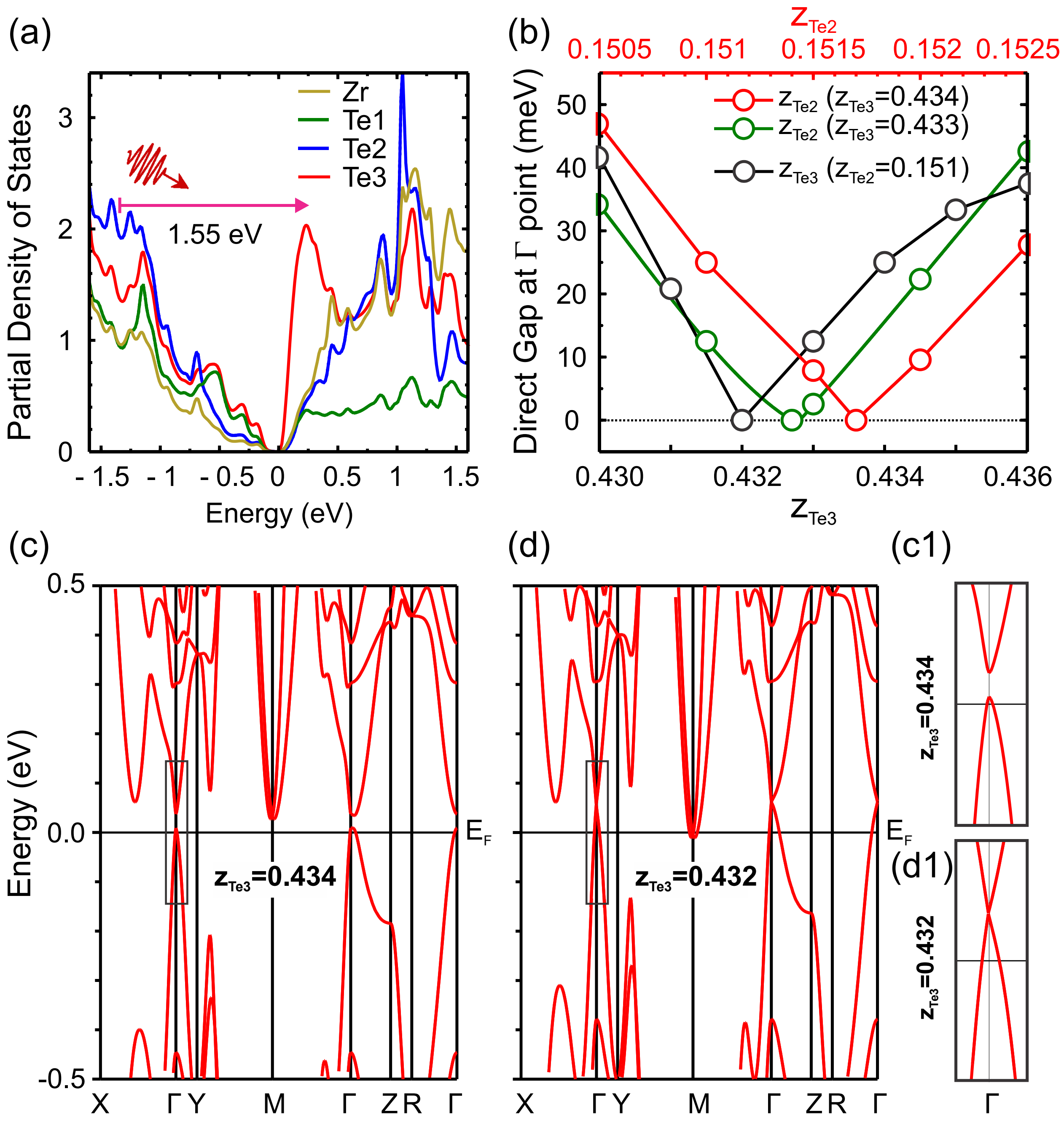

**Figure 4. The narrow-gap TI and Dirac semimetal electronic band structure of $ZrTe_5$ before and after photoexcitation calculated by DFT.** (a) The density of states (DOS) at time zero, for Te3 position $z_{\mathrm{Te3}} = 0.434$. The arrows illustrate the photoexcitation process where electrons from valence bands are excited to a Te3-dominated DOS peak near the Fermi level in the conduction band. (b) The direct gap at Γ point as a function of $z_{\mathrm{Te3}}$ position for $z_{\mathrm{Te2}} = 0.151$ (black) and of $z_{\mathrm{Te2}}$ position for $z_{\mathrm{Te3}} = 0.434$ (red) and $z_{\mathrm{Te3}} = 0.433$ (green). (c) The weak topological insulator band structure of $ZrTe_5$ in the equilibrium state at time zero ($z_{\mathrm{Te3}} = 0.434$); a small direct gap of ≈ 25 meV at the Γ point is clearly seen. (d) The band structure in the photoexcited state ($z_{\mathrm{Te3}} = 0.432$) has gap closed at the Dirac point, revealing a photoinduced phase transition from topological insulator to Dirac semimetal. (e) and (f) zoom in at the band structure near the Γ point (marked by the black rectangles) in panels (c) and (d), respectively.

# Supplementary Information for

## Photoinduced Dirac semimetal in $ZrTe_5$

T. Konstantinova, L. Wu, W.-G. Yin*, J. Tao, G. D. Gu, X.J. Wang, Jie Yang, I. A. Zaliznyak*, and Y. Zhu*

correspondence to: zaliznyak@bnl.gov, wyin@bnl.gov, zhu@bnl.gov,

**This PDF file includes:**

Supplementary Discussion
Supplementary Table 1
Supplementary Figures 1 to 5

**Supplementary Discussion**

Supplementary Table 1 shows atomic Wyckoff positions in $ZrTe_5$ which were used in our DFT calculations, including the $z_{\mathrm{Te3}}$ position before and after photoexcitation that was refined in our UED experiments.

| **Element** | **Wyckoff positions before photoexcitation** | | | | **New Wyckoff z-position after photoexcitation** |
|---|---|---|---|---|---|
| | | x | y | z | z |
| **Zr** | 4c | 0 | 0.316 | 0.250 | 0.250 |
| **Te1** | 4c | 0 | 0.663 | 0.250 | 0.250 |
| **Te2** | 8f | 0 | 0.993 | 0.151 | 0.15088 |
| **Te3** | 8f | 0 | 0.209 | 0.434 | 0.43211 |

**Supplementary Table 1**. Atomic Wyckoff positions in $ZrTe_5$ in the orthorhombic Cmcm structure [#63, lattice parameters $a$ = 3.9943Å, $b$ =14.547 Å and $c$ =13.749 Å at 300 K] before and after photoexcitation.

The imaginary part of the pseudo-dielectric function obtained from the ellipsometry measurements is presented in Supplementary Figure 1. The absorption peak of the inter-band transitions between the valence and the conduction bands is seen at ≈ 1.3 eV. This peak agrees very favorably with the DOS calculated by DFT (Fig. 4 of the main text and Supplementary Figure 5). There is significant absorption at 1.55 eV used for optical pumping in our experiments, which corresponds to the inter-band transitions indicated by horizontal arrow in Fig. 4 of the main text.

Supplementary Figure 2 presents the dependence of the ground state energy per formula unit and the direct band gap at the Γ point of $ZrTe_5$ on Te3 position. The energy minimum occurs for $z_{\mathrm{Te3}} = 0.434 - 0.435$, in excellent agreement with the experimentally determined equilibrium structure. In this structure, there is a small band gap of 25 – 35 meV. The ground state energy in the photoexcited state with $z_{\mathrm{Te3}} = 0.432$ is about 12 to 14 meV higher and the gap is zero, manifesting Dirac semimetal state.

The evolution of the DFT band structure of $ZrTe_5$ as a function of Te3 Wyckoff position for $z_{\mathrm{Te3}} = 0.430 - 0.437$ discussed in the main text is presented in Supplementary Figure 3. The vertical blue bars indicate the direct band gap at Γ point shown in Supplementary Figure 2 and its position with respect to Fermi energy, $E_F$.

The band structure and (partial) density of states (DOS) in $ZrTe_5$ structure at equilibrium and after photoexcitation presented in Supplementary Figure 4 show DFT results obtained without account for spin-orbit coupling (SOC). Both band structure and DOS are starkly different from results obtained when SOC is included (Supplementary Figure 5 and Fig 4 of the main text). This difference is especially marked near the Fermi level where without SOC a sizeable band gap exists both at equilibrium and in the photoexcited state. Account for SOC significantly modifies the electronic states, leading to the formation of chiral, spin-orbit-coupled band structure which is responsible for the unique unusual properties of $ZrTe_5$. Modification of such SOC-controlled bands involves long time scales associated with spin-orbit relaxation, which are observed in our UED experiments.

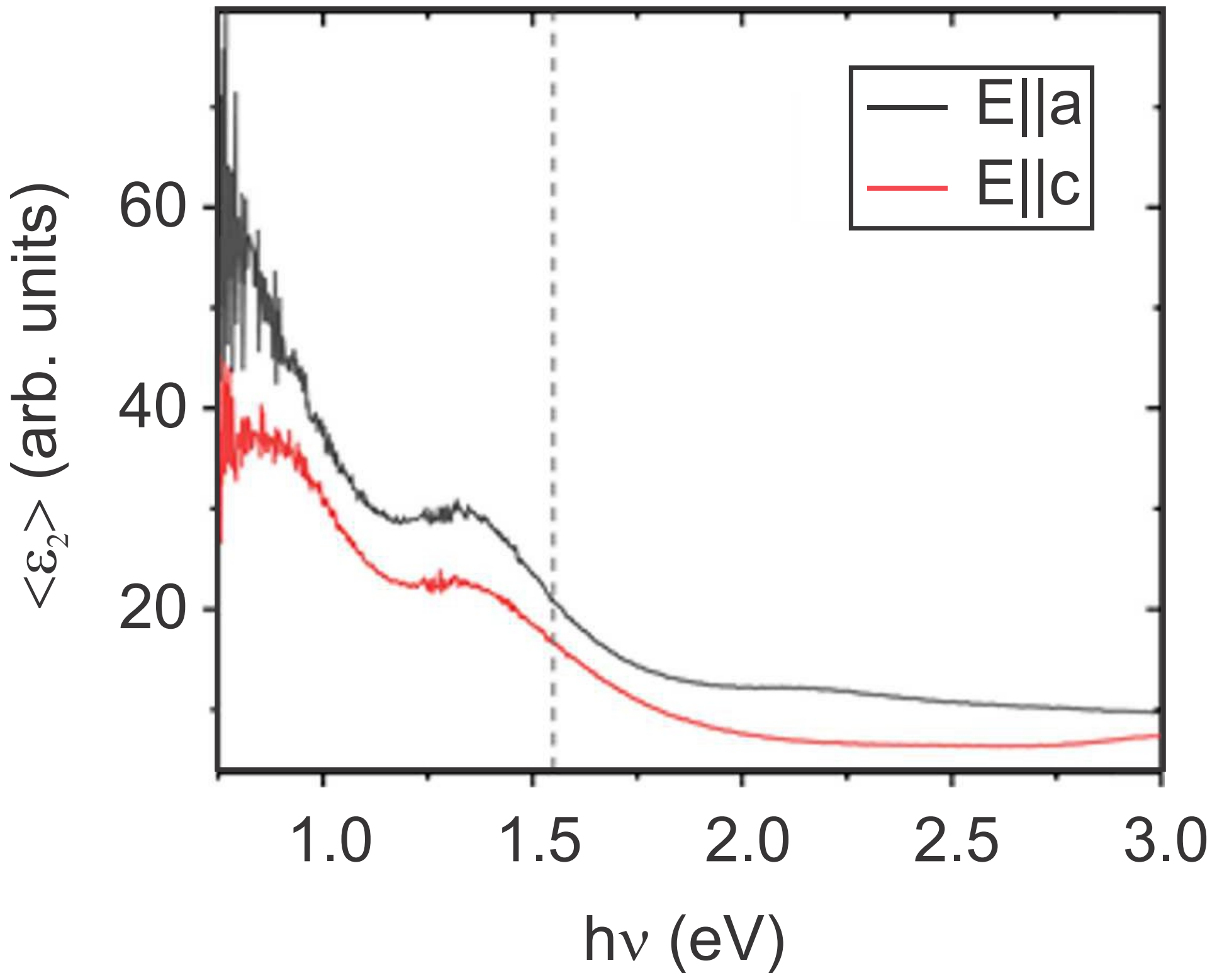

**Supplementary Figure 1.** $\langle \varepsilon_2 \rangle$ component of the pseudo dielectric function of $ZrTe_5$ with the light polarized along the $c$-axis (red) and $a$-axis (black). The dashed vertical line indicates the pump laser energy, $h\nu = 1.55$ eV.

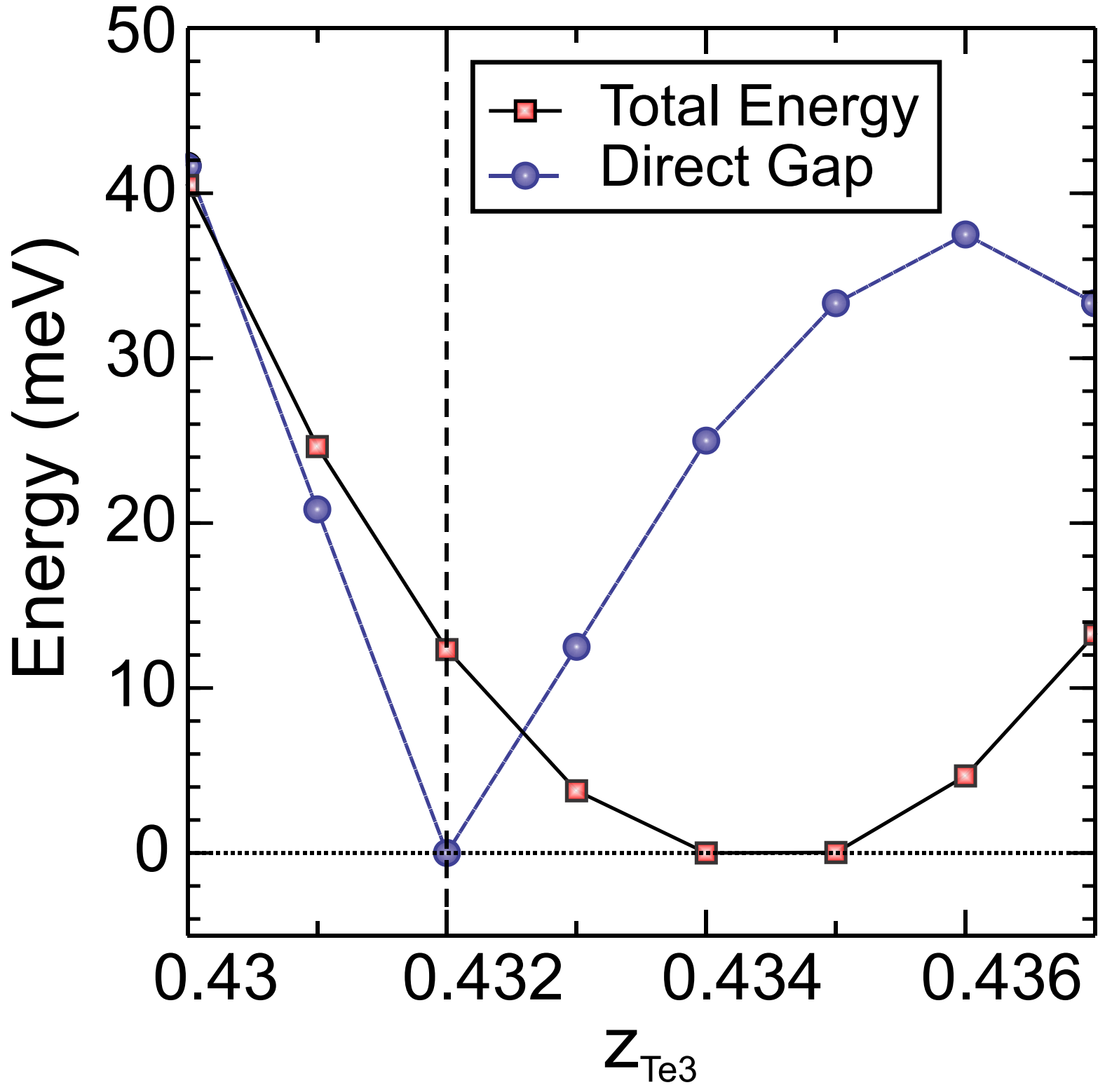

**Supplementary Figure 2.** The dependence of the direct gap at the Γ point (circles) and the ground state (GS) energy per formula unit of $ZrTe_5$ (squares) on Te3 Wyckoff position, $z_{\mathrm{Te3}}$, calculated by DFT. The gap is closed for $z_{\mathrm{Te3}} = 0.432$ (marked by vertical dashed line) where the GS energy is approximately 12 meV higher than the equilibrium GS energy before photoexcitation (for $z_{\mathrm{Te3}} = 0.434$), which is taken as zero (horizontal dotted line).

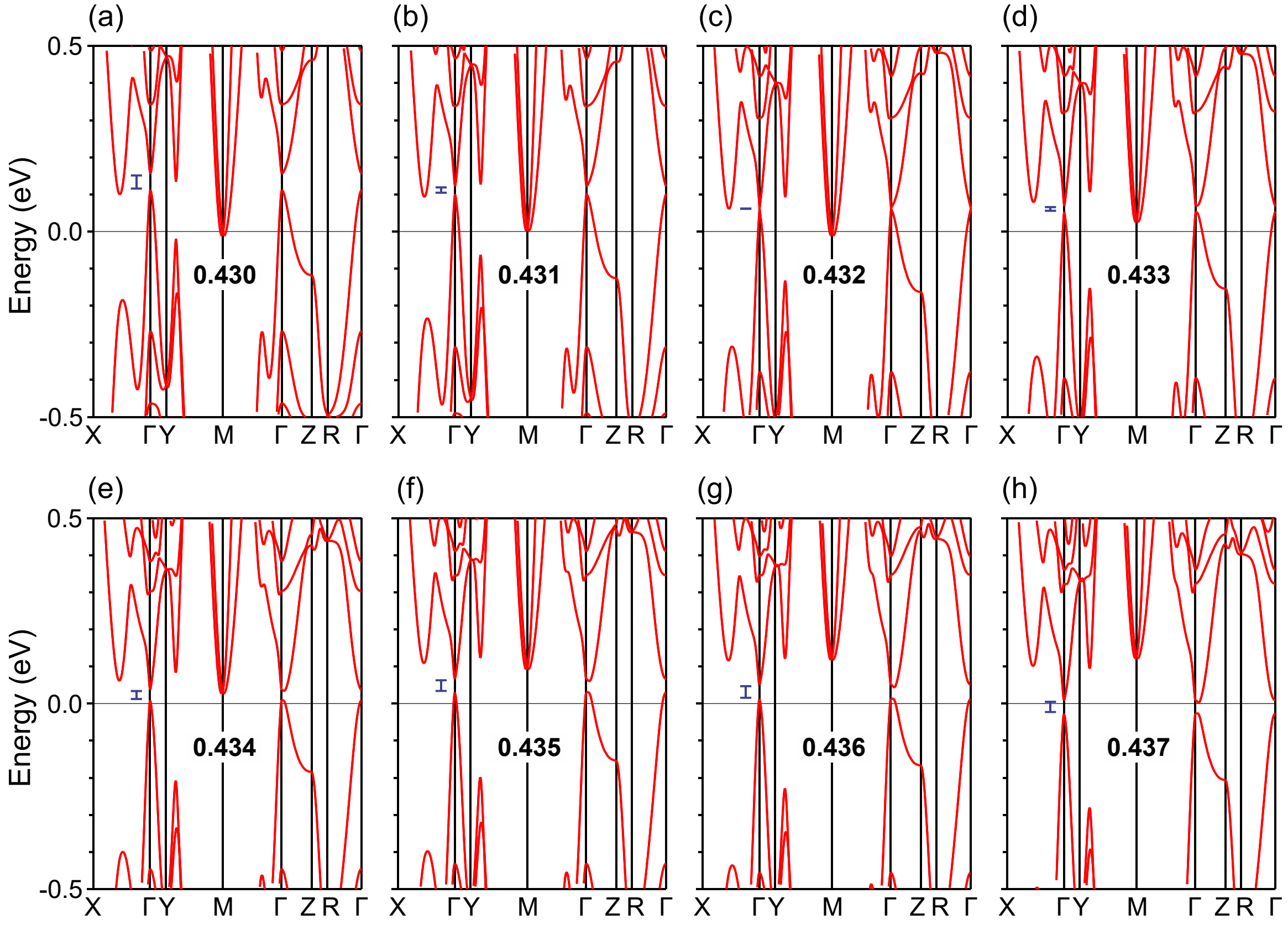

**Supplementary Figure 3.** The evolution of DFT band structure of $ZrTe_5$ as a function of Te3 Wyckoff position for $z_{\mathrm{Te3}} = 0.430 - 0.437$, (a)-(h), at $z_{\mathrm{Te2}} = 0.151$. The vertical blue bars illustrate size and position of the band gap at the Γ point. The gap is closed for $z_{\mathrm{Te3}} = 0.432$, which is the position refined from UED in the photoexcited state. The Dirac point emerges slightly (≈ 62 meV) above the Fermi energy, which hinders its direct observation by ARPES.

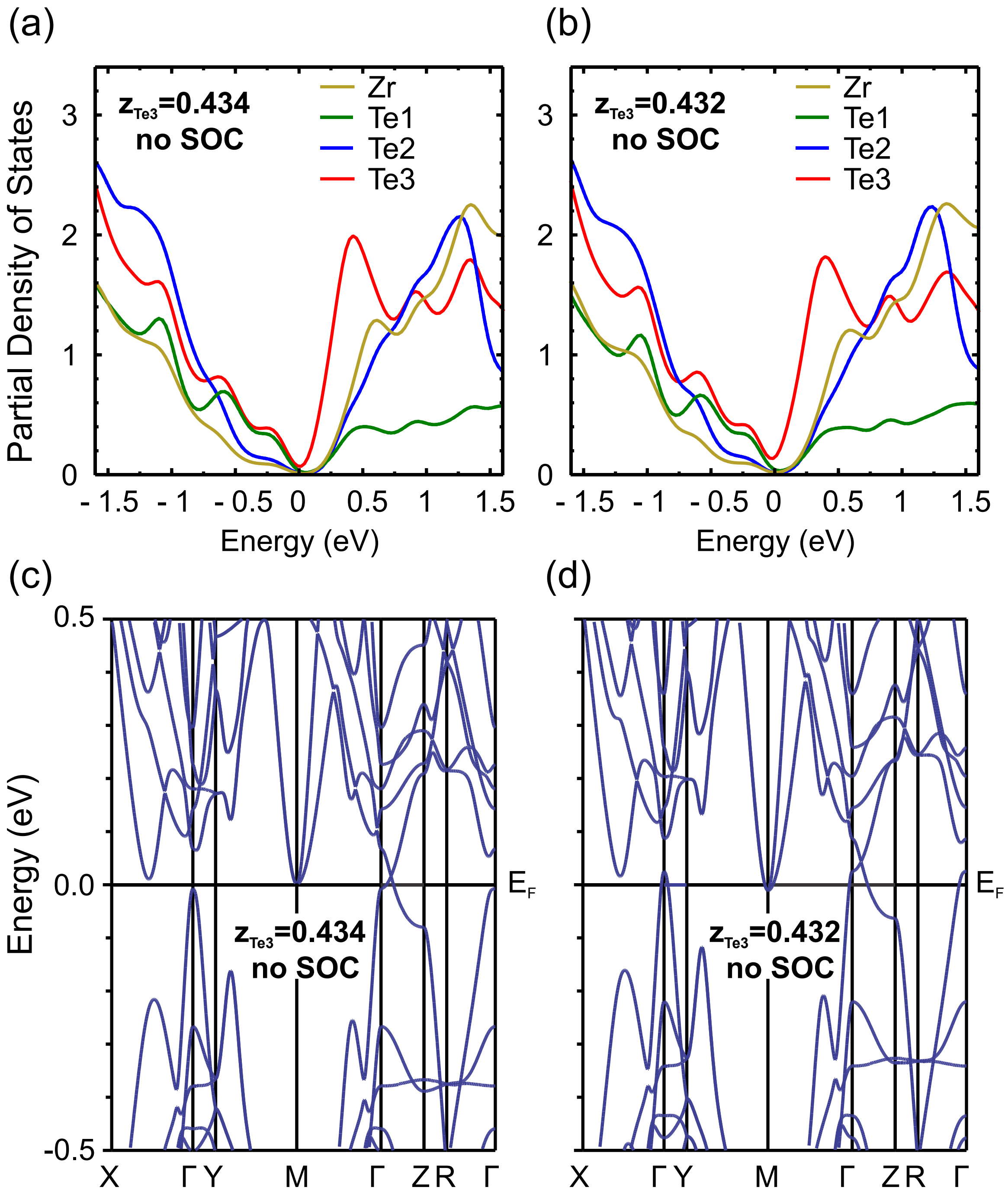

**Supplementary Figure 4.** The calculated DFT (partial) density of states and the corresponding electronic band structure of $ZrTe_5$ before (a), (c) and after (b), (d) photoexcitation in the absence of spin-orbital coupling (SOC), indicating no sign of photoinduced Dirac semimetal. The band structure is modified very significantly upon account of SOC (see Supplementary Figure 5).

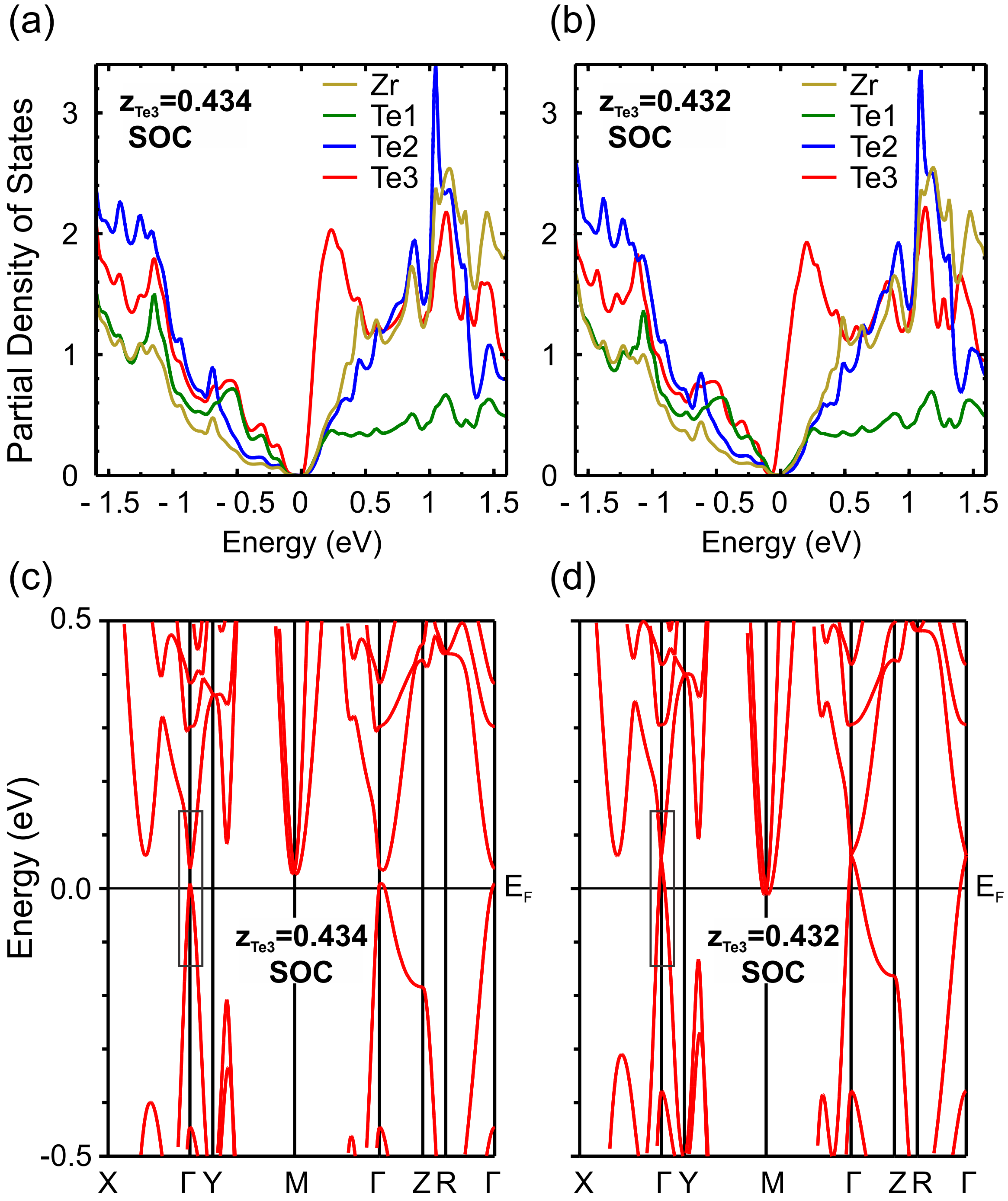

**Supplementary Figure 5.** The calculated DFT (partial) density of states and the corresponding electronic band structure of $ZrTe_5$ before (a), (c) and after (b), (d) photoexcitation with account for SOC, showing the emergence of photoinduced Dirac semimetal. The band structure is modified very significantly upon account of SOC.